\documentstyle[twocolumn,eclepsf]{jpsj}

\def\runtitle{
}
\def\runauthor{
}

\title{
Magnetic and Critical Properties of \\
Alternating Spin Heisenberg Chain in a Magnetic Field
}

\author{Tetsuji Kuramoto{\footnote{ kuramoto@physics.sci.ynu.ac.jp.}} 
}

\inst{
 Department of  Physics, Faculty of Engineering,
Yokohama National University, Hodogaya, Yokohama 240, Japan  \\
}

\recdate
{\today}

\abst{
We study  magnetic and critical properties of the alternating 
spin antiferromagnetic Heisenberg chain with $S=1/2$ and  $1$ in a magnetic field at $T=0$. The numerical diagonalization is applied to the 
system up to $2N=20$ sites. 
Checking numerically that  magnetic states with the magnetization per site $m$ obey a conformal field theory with conformal anomaly $c=1$ 
for $1/4<m<3/4$, we use the finite-size scaling of the conformal invariance to 
obtain a magentization curve in the thermodynamic limit.
In the magnetizatin curve a plateau appears at $m=1/4$. We also calculate 
two critical exponents $\eta$ and $\eta^z$ for $1/4<m<3/4$, 
which control the asymptotic behavior of the transverse and parallel spin correlation functions. We check the relation $\eta \eta^z=1$, which universally
 holds for a $c=1$ conformal field theory.   
}
\kword
{alternating spin chain, numerical diagnalization, 
conformal field theory}

\begin{document}
\sloppy
\maketitle
Recently, spin chains with an alternating array of different
  spins have attracted some current interest. Theoretically, Lieb and Mattis 
showed that the alternating spin antiferromagnetic Heisenberg chains with some 
next-nearest-neighour interactions have a ground state with large total spin.\cite{LM}
Very recently, numerical calculations of numerical diagonalization,
\cite{AL,ONO} density-matrix renomalization method\cite{DRG}  
and quantum Monte Carlo method \cite{MOTE} have been applied to the alternating spin 
Heisenberg chains. Experimentally, the alternating spin chains have been found 
as quasi-one-dimensional ferrimagnetic chains.\cite{EXP} 
On the other hand, the integrable alternating spin models have been constructed and solved 
via Bethe ansatz.\cite{WOY}  
Although its Hamiltonian has complicated nearest- and next-nearest-neighbour 
interactions, the integrable model has a single ground state.

Generalizing the Lieb-Shultz-Mattis theorem,\cite{LSM} 
Oshikawa et al gave a presence  condition of a magnetic plateau 
for spin chains in a magnetic field.\cite{OSHI} 
 The magnetic plateau has been observed in some spin models; for examples,
 an $S=1/2$ antiferromagnetic chain with period 3 exchange coupling\cite{HIDA} 
and  an $S=1$ antiferromagnetic chain with alternating bond.\cite{TONE}
The alternating spin chain is one of simple examples for the magnetic plateau.
Sakai and Takahashi performed the numercal calculation for 
$S=1$ antiferromagnetic Heisenberg chain in a magnetic field and
 revealed its magnetic and critical properties in the thermodynamic limit 
by using the technique of the conformal field theory.\cite{ST} 
Applying the Bethe ansatz to the integrable altenating spin chain 
with $S=1/2$ and $1$, Fujii et al calculated a magnetization  
curve and some critical exponents.\cite{FUJI} They did not observe any magnetic plateau but  a strange cusp in the magnetization curve. In this paper, following the procedure 
developed in ref.12, we study the magnetic and critical properties of the alternating 
spin antiferromagnetic Heisenberg chain with $S=1/2$ and  $1$ in a magnetic field at $T=0$ by numerical diagnoalization and use the finite-size 
scaling of the confomal field theory to find out its magnetic and critical properties in the themodynamic limit. 

 The alternating spin chain of $2N$ sites consists of $N$ spins
 $\mbox{\boldmath $\sigma$}_{1}$,$\mbox{\boldmath $\sigma$}_{3}$,$\cdots$,
$\mbox{\boldmath $\sigma$}_{2N-1}$ 
of spin $1/2$ and $N$ spins $\mbox{\boldmath $S$}_{2}$,  
$\mbox{\boldmath $S$}_{4}$, $\cdots$, $\mbox{\boldmath $S$}_{2N}$ of spin $1$.
This spin chain in a magnetic field $H$ is described by the Hamiltonian
\begin{eqnarray}
{\cal H} & = & {\cal H}_0 +{\cal H}_1
\label{hamil}
\end{eqnarray}
\begin{eqnarray}
{\cal H}_0 & = &
  \sum_{i=1}^{N} 
\left(
   \mbox{\boldmath $\sigma$}_{2i-1} \cdot \mbox{\boldmath $S$}_{2i}
    + \mbox{\boldmath $S$}_{2i} \cdot  \mbox{\boldmath $\sigma$}_{2i+1}
\right)
\label{hamil0}
\end{eqnarray}
\begin{eqnarray}
{\cal H}_1 & = &
 - H \sum_{i=1}^{N} 
\left(
       \sigma_{2i-1}^z + S_{2i}^z 
\right)
\label{hamil1}
\end{eqnarray}
where  we imopse periodic boundary conditions;
$\mbox{\boldmath $\sigma$}_{2N+1} = \mbox{\boldmath $\sigma$}_{1} $
and
$\mbox{\boldmath $S$}_{2N+2} =
\mbox{\boldmath $S$}_{2} $.
The Hamiltonian $\cal H$ is invariant under two-site translation  and 
rotation about $z$ axis. Thus,
all eigenstates of the Hamiltonian $\cal H$ can be classified by 
 the magnetization $M=\sum_{i=1}^N \left( \sigma_{2i-1}^z 
+ S_{2i}^z \right)$ and
  the wave vector $k$ ( $k= 2 \pi n/N $, $n=0,1,2, \cdots, N-1$ ).
We calculate the lowest energy $ E_k (2N,M)$ of the Hamiltonian ${\cal H}_0$   
in the subspace where the system has the magnetization $M$ and 
the wave vector $k$, by the numerical diagonalization with Lanczos algorithm
 up to $2N=20$.
We define $ E(2N,M)$ as the lowest energy of the Hamiltonian ${\cal H}_0$ 
 in the subspace with the magnetization $M$. In our case, we find that 
$E(2N,M)=E_{k=0}(2N,M)$.

The conformal field theory predicts that if the lowest-energy state 
with the magnetization per site $m$ is massless, 
the size-dependence of the lowest energy per site has the form\cite{CFT1}
\begin{eqnarray}
{\frac {1}{2N}} E(2N,M) \simeq \varepsilon (m) -{\frac {\pi}{6}}c& v_s&(m) {\frac {1}{(2N)^2}}\\
 &&(2N \rightarrow \infty)\nonumber
\label{scal0}
\end{eqnarray}
where we must vary $2N$ and $M$ with the magnetization per site fiexd at 
$m=M/(2N)$. $\varepsilon (m)$ is the lowest energy per site and 
$v_s(m)$ is the sound velocity. They depend on the magnetization per site $m$.
$c$ is the conformal anomaly.
\begin{figure}
\vspace*{0.5cm}
\begin{center}
\epsfile{file=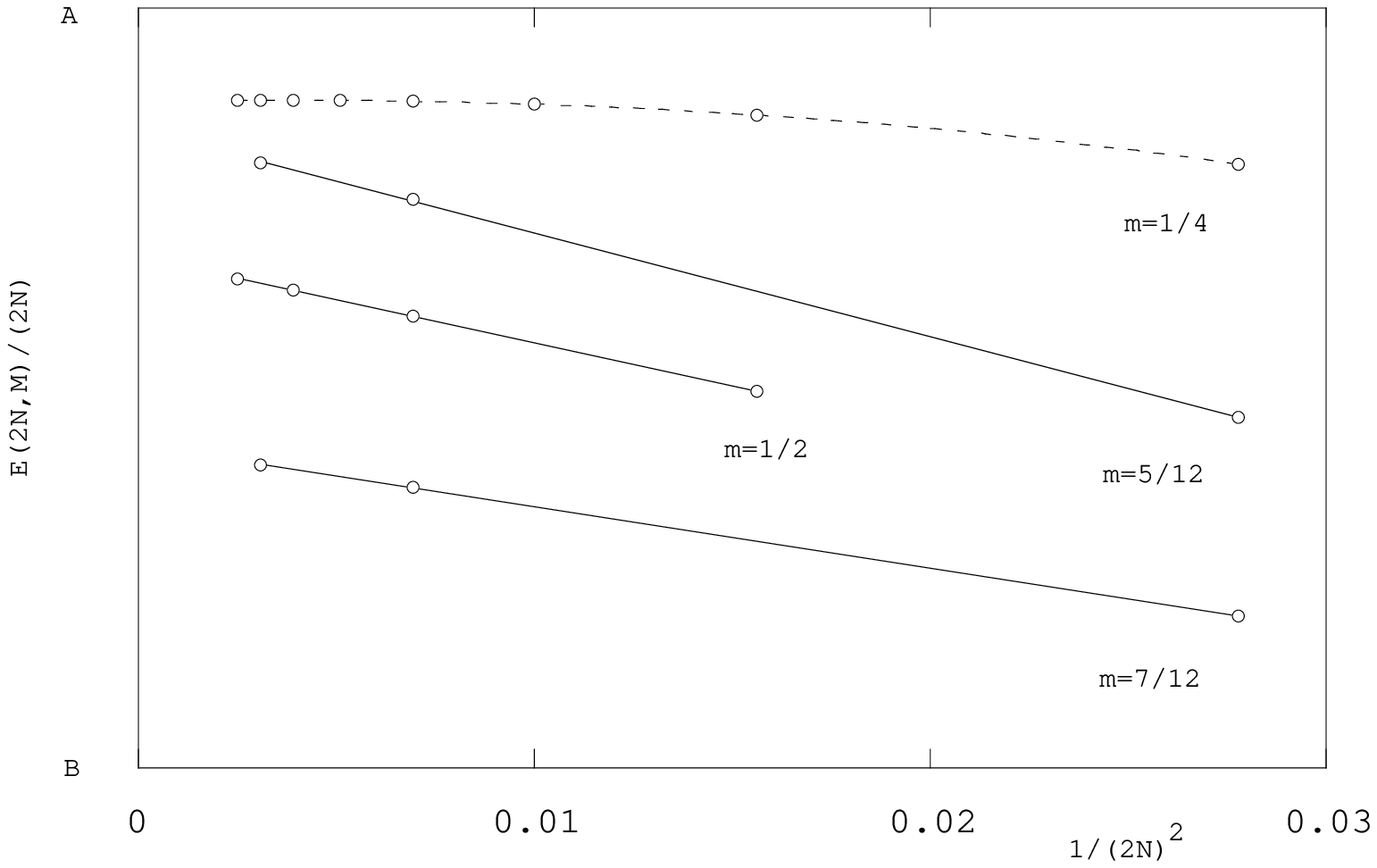,width=8.0cm}
\end{center}
\caption{Plots of $E(2N,M)/(2N)$ vs $1/(2N)^2$ with $m=M/(2N)$ fixed at 
$m=1/4$, $5/12$, $1/2$ and $7/12$. 
The origin is shifted along the vertical axis without changing the scale. 
The values of points A and B are as follows; $[A, B]=[-0.721, -0.771]$, 
$[-0.400, -0.450]$, $[-0.195, -0.245]$ and $[-0.040, -0.090]$ 
for $m=1/4$, $5/12$, $1/2$ and $7/12$, respectively. The solid lines show the fitting of two values 
$E(2N,M)/(2N)$ for the largest and next-largest $2N$ to the form (\ref{scal0}). 
 The dotted line is guide to the eye. }
\label{GE}
\end{figure}

 In Fig.\ref{GE}, we show plots of $E(2N,M)/{(2N)}$ versus 
$1/{(2N)^2}$ for $m=1/4$, $5/12$, $1/2$, $7/12$.
The plots are alomst linear for $m=5/12$, $1/2$, $7/12$
 but values for $m=1/4$ converge fast than $1/(2N)^2$. 
 Since the ground state of the alternating spin chain has the total spin $N/2$ 
and there is an energy gap $\Delta$ between the ground state and 
the lowest-energy state of the total spin $N/2+1$,\cite{LM} 
they suggets that the lowest energy state is massless for 
$1/4< m < 3/4$ but massive for  $0 \leq m \leq 1/4$.
 ( We use the energy gap $\Delta=1.767 \pm 0.003$ obtained in ref.5.)
 In the following, assuming that the lowest-energy state is massless for 
${1/4} < m < 3/4$, we apply the scaling law of conformal invariance 
to find out magnetic and critical properties in the thermodynamic limit.
To determine the conformal anomaly $c$, we need two values, the sound velocity
 $v_s$ and the gradient of the plot $E(2N,M)/{(2N)}$ versus $1/{(2N)^2}$, 
$ - \pi c v_s(m)/6$. To estimate the gradient,
we use two values of $E(2N,M)$ for the largest and next-largest $2N$ 
such that the magnetization per site is fixed at $m=M/(2N)$, 
up to $2N=20$. The next-lowest-energy state appears at the wave vector 
$k=2\pi/N$. The  gap between the lowest- and 
 next-lowest energies, $E_{2\pi/N}(2N,M)-E(2N,M)$,  has the dependence 
on the system size as\cite{AL,CFT1}
\begin{eqnarray}
 E_{2\pi/N}(2N,M)- E(2N,M) \simeq {\frac {\pi}{N}}& v_s&(m)\\
 &&(2N \rightarrow \infty).\nonumber
\label{scal1}
\end{eqnarray}
Thus, the sound velocity $v_s(m)$ is estimated by
\begin{eqnarray}
v_s(m)= {\frac{N}{\pi}}\left( E_{2\pi/N}(2N,M)-E(2N,M)\right)
\label{soundv}
\end{eqnarray}
for the largest $2N$. The results of the estimation for the conformal 
anomaly $c$  is shown in Table \ref{anomaly}. The value of $c$ for  $m=7/20$,
$3/8$ and $5/12$ is slightly larger than 1. 
\begin{fulltable}
\begin{center}
\begin{fulltabular}{p{4cm}cccccccccc}
\cline{2-11}
&&&&&&&&&&\\
&$m$ & $7/20$ &$3/8$ & $5/12$ &
$9/20$ &$1/2$ & $11/20$ & $7/12$ &
 $5/8$ & $13/20$ \\ 
&&&&&&&&&&\\
\cline{2-11}
&&&&&&&&&&\\
&$c$ & $1.20$ & $1.23$ & $1.18$ & $1.07$ &
$1.07$ &$1.04$ & $1.08$  &
 $1.07$ & $1.04$ \\ 
&&&&&&&&&&\\
\cline{2-11}
\end{fulltabular}
\end{center}
\caption{ Conformal anomaly $c$ estimated from egs. (\ref{scal0}) 
and (\ref{soundv}).}
\label{anomaly}
\end{fulltable}
As we will see later,
the universal relation between two exponents $\eta$ and $\eta^z$ for a $c=1$ 
conformal field theory holds in our case. Thus we conclude that $c=1$ 
for $1/4<m<3/4$.

\begin{figure}
\vspace*{0.5cm}
\begin{center}
\epsfile{file=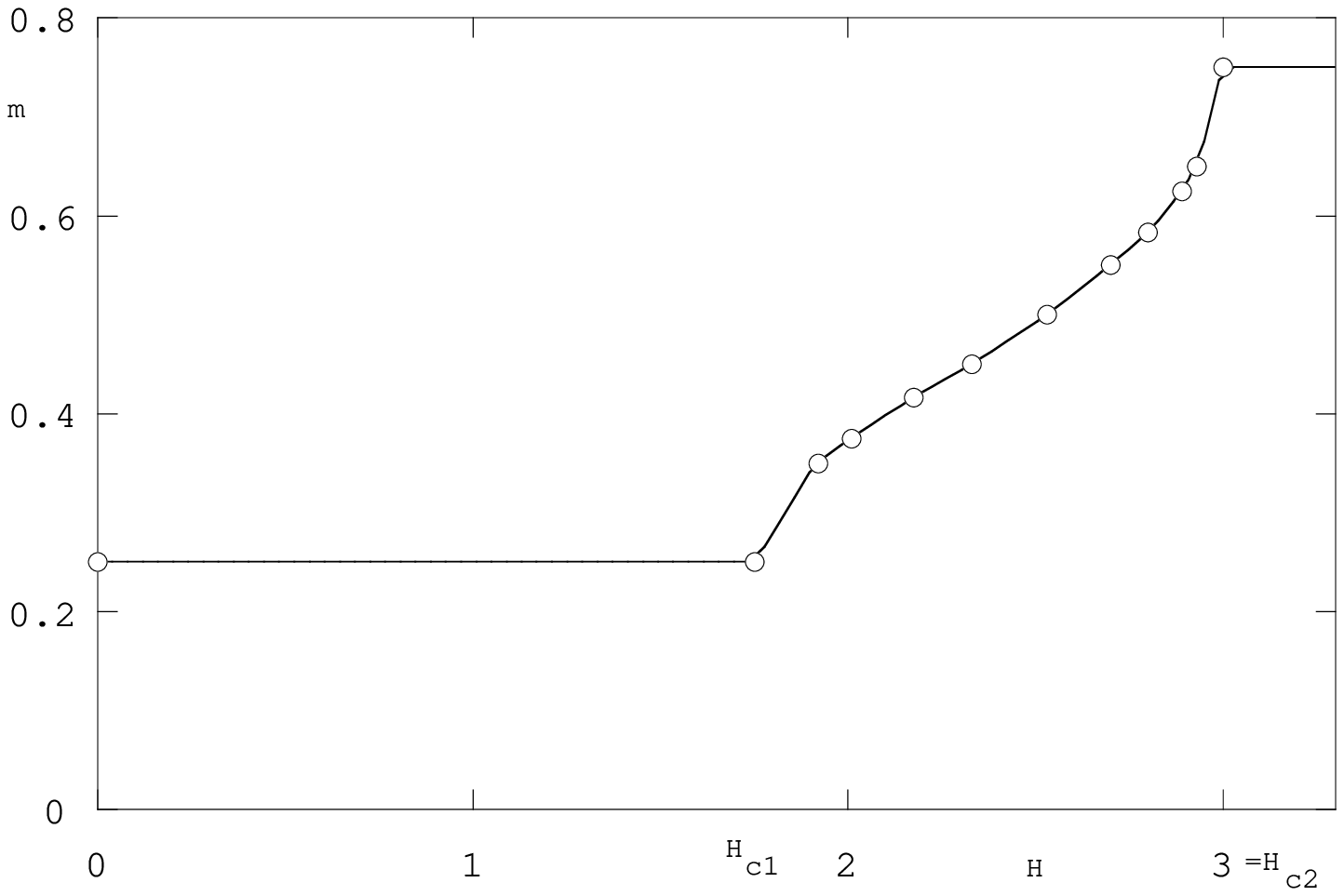,width=8.0cm}
\end{center}
\caption{Magnetization curve at the thermodynamic limit. Open circules 
 show plots of magnetization per site $m$ versus $H(=\varepsilon '(m))$ 
 which is estimated by averaging two results extrapolated 
by eqs. (\ref{spinE+}) and (\ref{spinE-}). A plateau appears at $m=1/4$. 
 }
\label{magC}
\end{figure}

We consider a magnetization curve of the alernating spin chain in two regions 
of $0<m<1/4$ and $1/4<m<3/4$. For $0<H<H_{c1}$, where the value $H_{c1}$ is 
equal to the energy gap $\Delta$, 
the lowest-energy state of the Hamiltonian $\cal H$ has the magnetization 
$M=N/2$. This means that $m=1/4$ for $0<H<H_{c1}(=\Delta)$.
For $1/4<m<3/4$, owing to the confomal invariance, 
the spin-excitation energy of the Hamiltonian ${\cal H}$ 
depends on the system size $2N$ as
\begin{eqnarray}
E(2N,M+1)-E(2N,M)-H \simeq \pi v_s \eta {\frac{1}{2N}}  
\label{spinE+}
\end{eqnarray}
\begin{eqnarray}
E(2N,M)-E(2N,M-1)-H \simeq - \pi v_s \eta {\frac{1}{2N}}  
\label{spinE-}
\end{eqnarray}
in the limit $2N \rightarrow \infty$,\cite{CFT1} 
where we vary $2N$ and $M$ with $m=M/(2N)$ fixed. 
$\eta$ is a critical exponent of correlation 
function for spin-excitations, which we will estimate in the following.
\begin{figure}
\vspace*{0.5cm}
\begin{center}
\epsfile{file=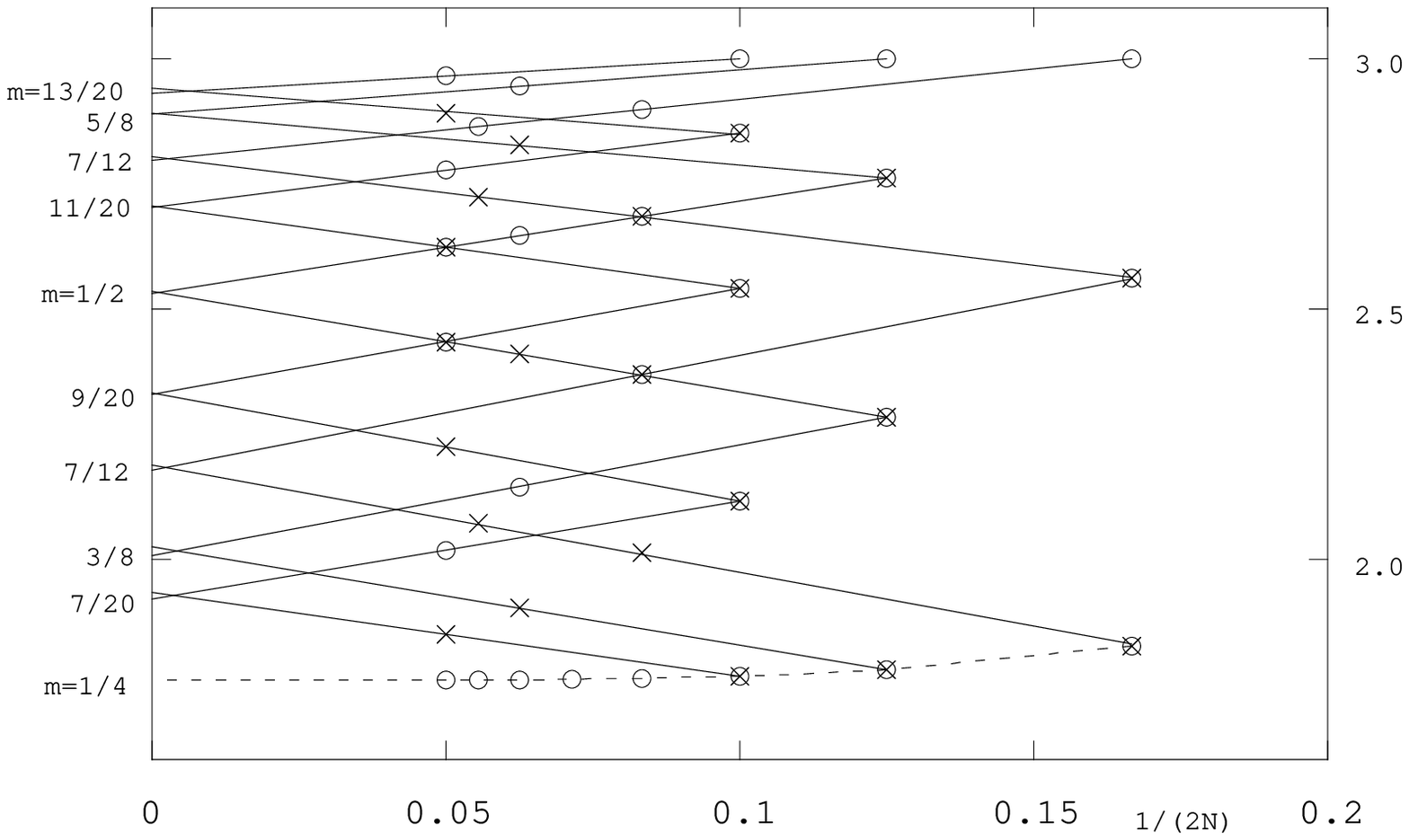,width=8.0cm}
\end{center}
\caption{Plots versus $1/(2N)$ of spin-excitation gaps, $E(2N,M+1)-E(2N,M)$ (open circles) and $E(2N,M)-E(2N,M-1)$ (crosses), 
with $m=M/(2N)$. The solid lines show the fitting of two values of 
 $E(2N,M+1)-E(2N,M)$ ( $E(2N,M)-E(2N,M-1)$ ) for the largest and 
next-largest $2N$ to the form (\ref{spinE+}) ( (\ref{spinE-}) ). 
 For $m=1/4$, the gap $E(2N,M+1)-E(2N,M)$, connected by a dotted line, 
converges to a finite value $\Delta$. }
\label{SPINE}
\end{figure}
In Fig. \ref{SPINE} ,  we plot $E(2N,M+1)-E(2N,M)$ and $E(2N,M)-E(2N,M-1)$ versus $1/{(2N)}$. The plots are almost linear at least for $m=5/12$, $1/2$ 
and $7/12$. For $m=1/4$, the gap $E(2N,M+1)-E(2N,M)$ converges
faster than $1/(2N)$.  
Combinning eqs. (\ref{scal0}) and (\ref{spinE+}) (or (\ref{spinE-})) and taking 
the limit $2N \rightarrow \infty$ ,  we get the relation between
 the magentic field $H$ and magnetization  per site $m$, $\varepsilon '(m)=H$,
which gives us a magnetization curve at $T=0$. In the following, we estimate
$\varepsilon '(m)$ for several $m$'s and connect them by a smmoth curve. 
To etsimate  $\varepsilon '(m)$ for $1/4<m<3/4$, 
we use the largest- and next-largest-size data for $E(2N,M+1)-E(2N,M)$ 
in the form (\ref{spinE+}) and do the same treatment 
for $E(2N,M)-E(2N,M-1)$ in the form (\ref{spinE-}). Two results of 
$\varepsilon '(m)$ for $9/20\leq m$ coincide with each other 
within a difference less than $0.5 \% $. The difference for $m=7/20$,
 $3/8$ and $7/12$ is less than $1 \% $. Thus we regard 
the averge of the two results as the extraporated value of $\varepsilon '(m)$.
Using the extrapolated values of $\varepsilon '(m)$, we draw in Fig.\ref{magC} 
 the magnetization curve at $T=0$ in the thermodynamic limit.
We see that a plateau appears at $m=1/4$.
The presence of this plateau is expected from the Lieb-Mattis theorem,\cite{LM}
  which shows that the ground state of the alternating spin 
antiferromagnetic Heisenberg chain has the total spin $N/2$
 and has an energy gap between the ground state and the lowest-energy state 
of spin $N/2+1$. It is also consistent with the generalized LSM theorem 
proposed by Oshikawa et al \cite{OSHI}. The generalized LSM theorem can 
be applied to general spin chains with axial symmetry. 
The quantization condition of a magnetic plateau in our case is that ${1/2}+1-2m=$intger, which holds for $m=1/4$.

\begin{figure}
\vspace*{0.5cm}
\begin{center}
\epsfile{file=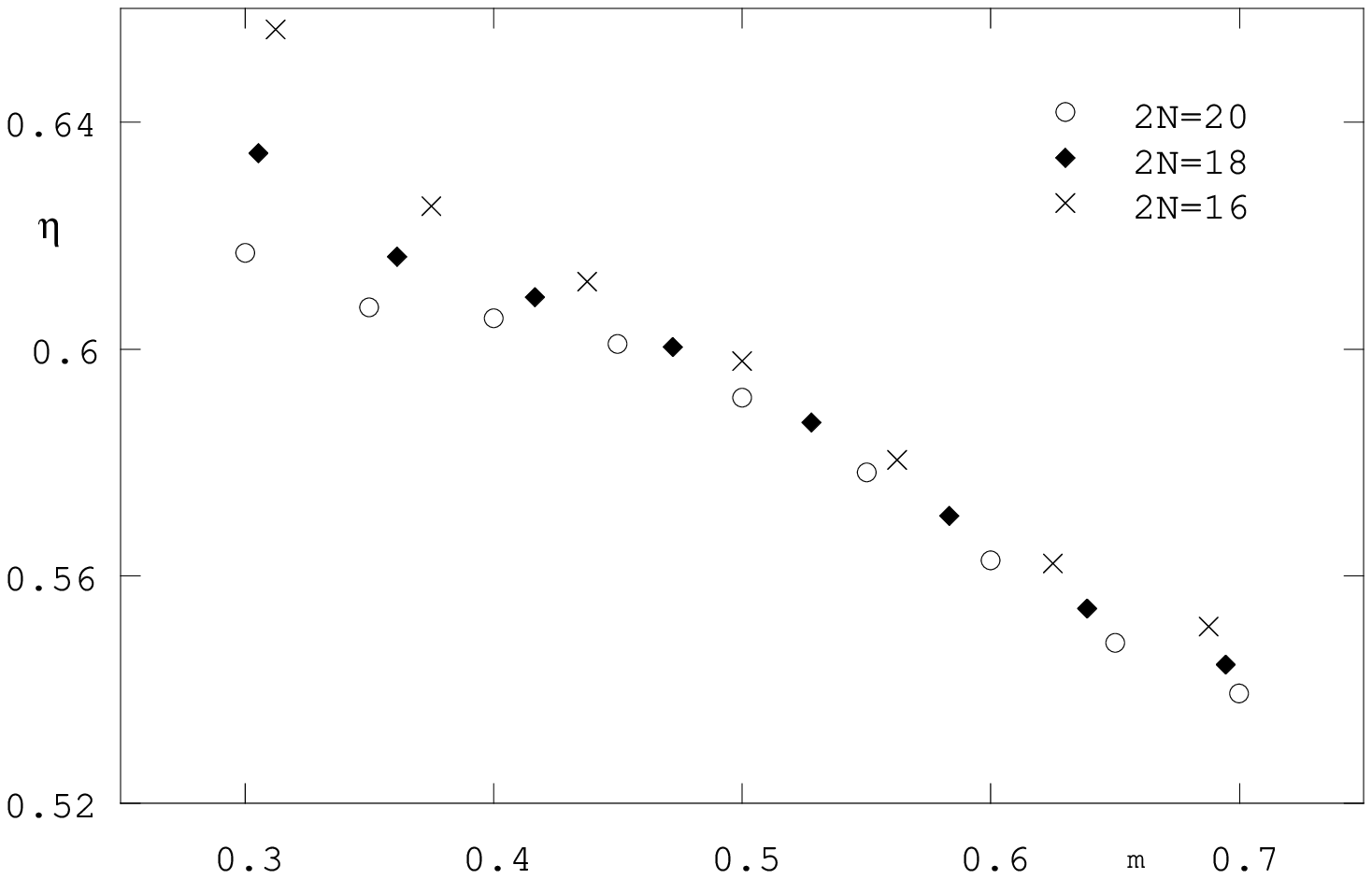,width=8.0cm}
\end{center}
\caption{
Critical exponent  $\eta$ estimated from eq.(\ref{eta}) for $2N=16$, $18$ 
and $20$.}
\label{expeta}
\end{figure}

Now we calculate two critical exponents, $\eta$ and $\eta ^z$, 
for correlation functions of the transverse and parallel components 
to a magnetic field, which behave as $(-1)^r r^{-\eta} $ for the transverse component
 and $ r^{-\eta^z}{\rm cos}(2k_F r)$ for the parallel one, where $r$ is the distance between two operators in the correlation function 
and $2k_F$ is a wave vector whose definition will be given later.  
First we consider the critical exponent $\eta$.
The exponent $\eta$ appears in the asymptotic forms (\ref{spinE+}) and
(\ref{spinE-}) of the spin-excitation energy gaps.  
Thus, substitutting eq.(\ref{spinE-}) from eq.(\ref{spinE+}) 
and using eq.(\ref{soundv}), 
we get the relation 
\begin{eqnarray}
\eta&=&\nonumber\\
& &{\frac{E(2N,M+1)+E(2N,M-1)-2 E(2N,M)} {E_{2\pi/N}(2N,M)- E(2N,M)}} \nonumber\\ 
\label{eta}
\end{eqnarray}
 The results of $\eta$ evaluated by eq. (\ref{eta}) for $2N=16$, $18$ and $20$ are plotted in Fig.\ref{expeta} versus magnetization per site $m$.
We determine $\eta$ in the limits of 
$m\rightarrow (1/4)\,+$ and  $m\rightarrow (3/4)\,-$, by extrapolating 
the value of $\eta$ for $M=N/2+1$ $(m=(1/4)\,+)$ and $M=3N/2-1$ $(m=(3/4)\,-)$ 
linearly to $1/(2N)$ for $2N=18$ and $20$. 
The results are $\eta=0.460$ for $m=1/4$ and $\eta=0.493$ for $m=3/4$.

\begin{figure}
\vspace*{0.5cm}
\begin{center}
\epsfile{file=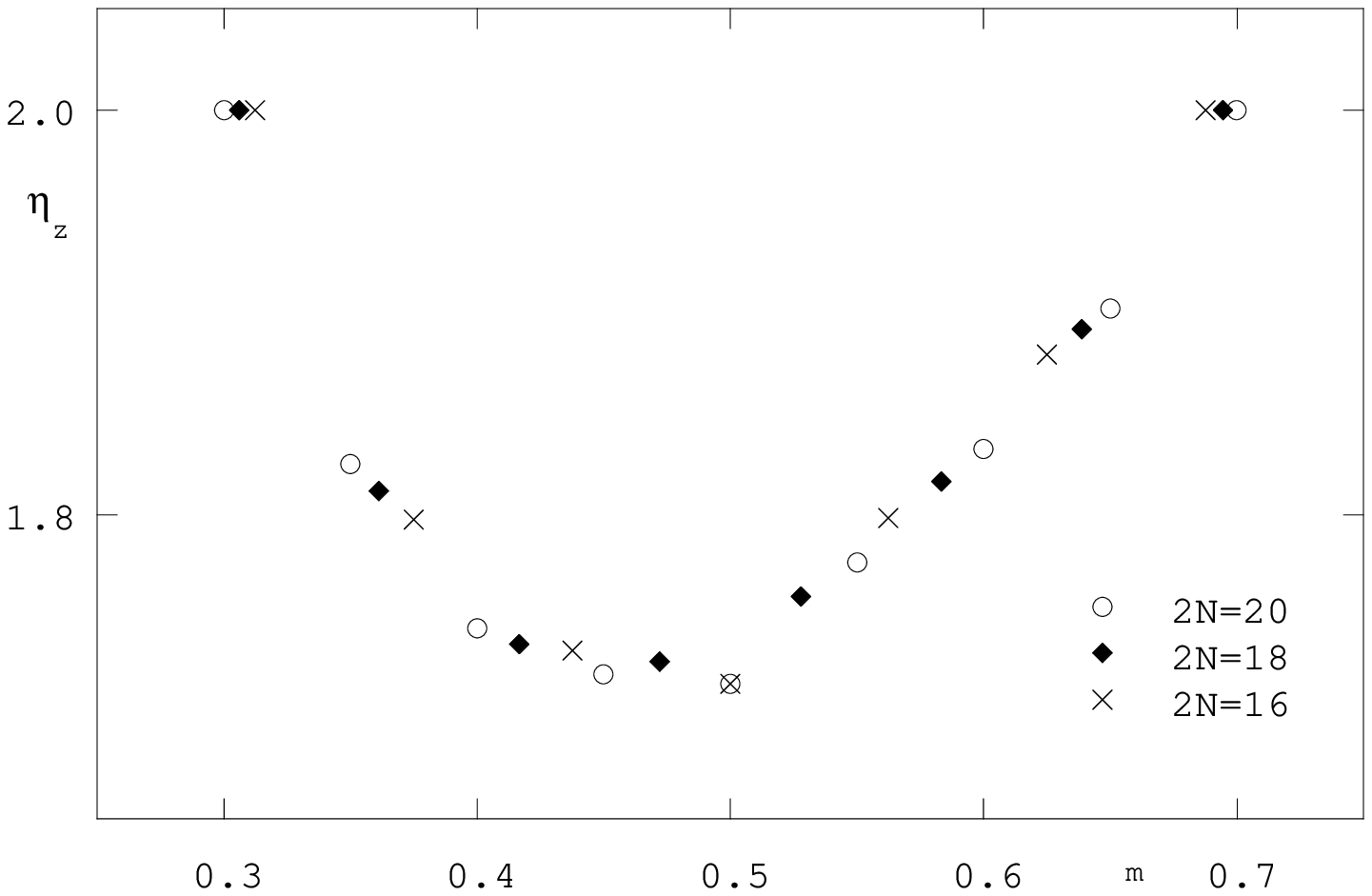,width=8.0cm}
\end{center}
\caption{Critical exponent  $\eta^z$ estimated from eq. (\ref{etaz}) 
for $2N=16$, $18$ and $20$.}
\label{expetaz}
\end{figure}

Next we consider the critical exponent $\eta^z$ for the spin-spin correlation function.
  In the numerical calculation of $E_k (2N,M)$ for $k=0$, $\pi/N$, $2\pi/N$,$\cdots$, 
$2\pi (1-1/N)$, we find that a soft mode exits 
at $k=2\pi M/N-\pi \equiv 2 k_F$ for $M\leq N$ and 
 $k=3\pi-2\pi M/N\equiv 2 k_F$ for $M < N$ . 
If the soft mode become a gapless excitation as $2N\rightarrow \infty$, 
the conformal invariance predicts that its energy gap from the lowest energy depends 
on the system size $2N$ as
\begin{eqnarray}
E_{2k_F}(2N,M)-E(2N,M)\simeq\pi v_s \eta^z\frac{1}{2N} 
\label{spin-spinE}
\end{eqnarray}
in the limit $2N\rightarrow \infty$.\cite{CFT2} We  numerically check this asymptotical 
behavior of the energy gap 
 for $m=5/12$, $1/2$ and $7/12$. Thus we assume 
that the gapless excitation exits at $k=2k_F$ in the infinite system. 
 To estimate $\eta^z$, we use the relation 
\begin{eqnarray}
\eta^z=2 {\frac{E_{2k_F}(2N,M)-E(2N,M)} {E_{2 \pi/N}(2N,M)- E(2N,M)}}  
\label{etaz}
\end{eqnarray}
derived from eqs. (\ref{soundv}) and (\ref{spin-spinE}).
The results of $\eta^z$ for $2N=16$, $18$ and $20$ are plotted 
in Fig.\ref{expetaz}.

We compare the above results of $\eta$ and $\eta^z$ with those for 
$S=1/2$ and $S=1$ antiferromagnetic Heisenberg chains.
Let us first note the behaviors of $\eta$ and $\eta^z$ for $S=1/2$ 
and $S=1$.
For $S=1/2$, when $m$ varies from $0$ to $1/2$, 
$\eta$ monotonously decreases from $1$ to $1/2$ and 
$\eta^z$ monotonously increases from $1$ to $2$.\cite{CFT3}
For $S=1$, $\eta=1/2$ and $\eta^z=2$ at $m=0$ and $m=1$. 
$\eta$ and $\eta^z$ have a minimum and a maximum near $m=1/3$,
  respectively.\cite{ST}
The parallel spin correlation for $S=1/2$ is stronger than that for $S=1$.
In the alternating spin chain, $\eta$ and $\eta^z$ have a maximum
 and a minimum, respectively. The parallel spin correlation for the alternating 
spin chain is stronger than that for $S=1$ but the transverse one is weaker.
The alternating spin is similar in this aspect to $S=1/2$.

Finally, let us check that two estimated critical exponents $\eta$ and $\eta^z$ satisfy an univeral relation 
$\eta \eta^z=1$ for $1/4< m <3/4$. This relation holds universally for a $c=1$ conformal field theory.\cite{LUTT} The values of $\eta \eta^z$ 
for each $m$ in the case of $2N=20$ are shown in Table \ref{relation}. 
It shows that the relation $\eta \eta^z=1$ is satisfied within errors.

It is a pleasure to thank T. Ono for valuable discussions. 
The computation in this paper has been done partially at Supercomputer Center, 
ISSP, University of Tokyo.
\begin{table}
\begin{center}
\begin{tabular}{p{2cm}cccc}
\cline{2-5}
&&&&\\
&$M$ & $\eta$ &$\eta^z$ & $\eta \eta^z$  \\ 
&&&&\\
\cline{2-5}
&&&&\\
&$6$ & $0.617$ &$2.000$ & $1.234$  \\ 
&$7$ & $0.608$ &$1.825$ & $1.109$  \\ 
&$8$ & $0.606$ &$1.744$ & $1.056$  \\ 
&$9$ & $0.601$ &$1.721$ & $1.034$  \\ 
&$10$ & $0.592$ &$1.717$ & $1.016$  \\ 
&$11$ & $0.578$ &$1.777$ & $1.027$  \\ 
&$12$ & $0.563$ &$1.833$ & $1.031$  \\ 
&$13$ & $0.548$ &$1.902$ & $1.043$  \\ 
&$14$ & $0.539$ &$2.000$ & $1.078$  \\
&&&&\\ 
\cline{2-5}
\end{tabular}
\end{center}
\caption{ Exponents $\eta$ and $\eta^z$ estimated from eqs. (\ref{eta}) 
and (\ref{etaz}) for $2N=20$  and the values of $\eta \eta^z$. }
\label{relation}
\end{table}

%


\begin{thebibliography}{99}
%
\bibitem{LM} E. Lieb and D. Mattis: J. Math. Phys. {\bf 3} (1962) 749.
\bibitem{AL} F. C. Alcaraz and A. L. Malvezzi: To be published in J. Phys. A.
\bibitem{ONO} T. Ono, T. Nishimura, M. Katsumura, T. Morita and M. Sugimoto:
J. Phys. Soc. Jpn. {\bf 66} (1997) 2576.
\bibitem{DRG} S.K. Pati,S. Ramasesha and D.Sen: preprint, cond-mat /9704057
;H. Niggeman, G. Uimin and J. Zittartz: J. Phys. :Condens. Matter {\bf 9} 
(1997) 9031.
\bibitem{MOTE}S. Brehmer, H.-J. Mikeska and S. Yamamoto: 
    J. Phys.: Condens. Matter {\bf 9} (1997) 3921.
\bibitem{EXP}
M. Verdaguer, M. Julve, A. Michalowicz and O. Kahn : Inorg. Chem. {\bf 22} (1983) 2624; Y. Pei, M. Vergaguer, O. Kahn, J. Sletten and J.-P. Renard :
 Inorg. Chem. {\bf 26} (1987) 138; E. Coronado, M. Drillon, P. R. Nugteren, 
L. J. de Jongh, D. Beltran and R. Georges : J. Am. Chem. Soc. {\bf 111} 
(1989) 3874 ;   P. Zhou, B. G. Morin and A. J. Epstein: J. Appl. Phys. 
{\bf 73} (1993) 6569.
\bibitem{WOY} H. J. de Vega and F. Woynarovich:
 Nucl. Phys. B {\bf 251} (1985)  439
(1985); F. Woynarovich, J. Phys. A {\bf 22} (1989) 4243.
\bibitem{LSM}E. H. Lieb, T. Schultz and D. J. Mattis: Ann. Phys. (N. Y.) 
{\bf 16} (1961) 407;   I. Affleck and E. H. Lieb, Lett. Math. Phys. 
         {\bf 12} (1986) 57.
\bibitem{OSHI} M. Oshikawa, M. Yamanaka and I. Affleck: Phys. Rev. Lett. 
{\bf 78} (1997) 1984.
\bibitem{HIDA} K. Hida: Phys. Soc. Jpn. {\bf 63} (1994) 2359.
\bibitem{TONE} T. Tonegawa, T. Nakano and M. Kaburagi: 
   J. Phys. Soc. Jpn. {\bf 65} (1996) 3317.
\bibitem{ST} T. Sakai and M. Takahashi: Phys. Rev. B {\bf 43}(1991) 13383;
{\it ibid}: J. Phys. Soc. Jpn. {\bf 60} (1991) 3615.
\bibitem{FUJI} M. Fujii, S. Fujimoto and N. Kawakami: J. Phys. Soc. Jpn. 
                  {\bf 65} (1996) 2381. 
\bibitem{CFT1} 
J. L. Cardy: Nucl. Phys. B {\bf 270} [FS16] (1986) 186;
H. W. Br\"ote, J. L. Cardy and M. P. Nightingale:
 Phys. Rev. Lett.  {\bf 56} (1986) 742;
I. Affleck, Phys. Rev. Lett.  {\bf 56}  (1986) 746.
\bibitem{CFT2}N. M. Bogoliubov, A. G. Izergin and N. Yu Reshetikhin: J. Phys. 
A {\bf 20} (1987) 5361.
\bibitem{CFT3}N. M. Bogoliubov, A. G. Izergin and V. E. Korepin: Nucl. Phys. 
B {\bf 275} [FS17] (1986) 687.
\bibitem{LUTT}F. D. M. Haldane: Phys. Rev. Lett. {\bf 45} (1980) 1358;
 Phys. Lett. A {\bf 81} (1981) 153 ; J. Phys. C {\bf 14} (1981) 2585.

\end{thebibliography}
\end{document}